\documentclass{ws-ijmpa}

\begin{document}

\markboth{C.~Piskor-Ignatowicz for the COSY-11 Collaboration}
{}

%
\catchline{}{}{}{}{}
%

\title{ NEAR THRESHOLD $\eta$ MESON PRODUCTION IN $dp$ COLLISIONS
}

\author{\footnotesize C.~Piskor-Ignatowicz$^{\star}$$^,$\footnote{E-mail address: 
c.piskor-ignatowicz@fz-juelich.de}~, 
J.~Smyrski$^{\star}$,
P.~Moskal$^{\star}$, 
H.-H.~Adam$^{\#}$, 
A.~Budzanowski$^{\$}$,
E.~Czerwi\'nski$^{\star}$,
R.~Czy\.zykiewicz$^{\star}$,
D.~Gil$^{\star}$,
D.~Grzonka$^{\%}$, 
M.~Janusz$^{\star}$, 
L.~Jarczyk$^{\star}$, 
B.~Kamys$^{\star}$, 
A.~Khoukaz$^{\#}$, 
P.~Klaja$^{\star,\%}$, 
J.~Majewski$^{\star,\%}$, 
W.~Oelert$^{\%}$,   
J.~Przerwa$^{\star,\%}$, 
J.~Ritman$^{\%}$, 
T.~Ro\.zek$^{+}$, 
T.~Sefzick$^{\%}$, 
M.~Siemaszko$^{+}$,  
A.~T\"aschner$^{\#}$, 
P.~Winter$^{\times}$,  
M.~Wolke$^{\%}$, 
P.~W\"ustner$^{\%}$, 
W.~Zipper$^{+}$ 
}

\address{
$^{\star}$Institute of Physics, Jagiellonian University, Cracow, Poland\\ 
$^{\#}$IKP, Westf\"alische Wilhelms-Universit\"at, M\"unster, Germany\\
$^{\$}$Institute of Nuclear Physics, Cracow, Poland\\ 
$^{\%}$IKP, ZEL, Forschungszentrum Juelich, Juelich, Germany\\ 
$^{+}$Institute of Physics, University of Silesia, Katowice, Poland\\
$^{\times}$Department of Physics, University of Illinois at Urbana-Champaign, Urbana, IL 61801 USA\\
}

\maketitle


\begin{abstract}
Preliminary results of recent measurements of the near threshold $\eta$ meson production
in the $dp\rightarrow dp\eta$ reaction are presented. 
The experiment was performed at the COSY-Juelich accelerator with the use of the COSY-11 
detection system. Data were taken for three 
values of deuteron beam momenta corresponding
to excess energies of 3.2, 6.1 and 9.2~MeV. 
The energy dependence of the total cross section confirms a strong effect 
of the final state interaction.

\end{abstract}

\section{Introduction}  
The three nucleon system is the simplest system for study
of the multi-nucleon interaction and, in particular, of inelastic processes in nuclear matter.
From experimental point of view, the most suitable for these studies are the $pd$
(or $dp$) collisions due to the high quality of proton (deuteron) beams available at particle
accelerators. Recently the $d- \eta$ and  ${^3\mbox{He}}-\eta$ interaction was intensively studied on 
the theoretical ground. This interaction is of special interest due to the possible existence of 
the $\eta$-nucleus bound or quasi-bound states. From the experimental point of view, 
the $\eta$ meson production near threshold in the three nucleon system is much less explored 
as compared to the two nucleon system\cite{moskal02,hanha04,faldt02}.
The database for the $pd\to {^3\mbox{He}}\eta$ 
reaction\cite{mayer96,betigeri,bilger02,hha2,smy06} has improved recently, however,
for the reaction $pd\rightarrow dp\eta$ close to threshold there exist only data
measured with the SPESIII spectrometer at SATURNE\cite{Hibou00}
for two excess energies namely $Q$ = 1.1 and 3.3~MeV.
Unfortunately, the uncertainty of the excess energy of theses data points
of $\Delta Q = \pm 0.6$~MeV is very large when taking into account the rapid rise
of the cross section near threshold.  
At higher energies, for $Q \ge$  14~MeV, the total and differential 
distributions of the cross section were determined 
by the PROMICE/WASA collaboration\cite{zlo03,bilg04}.
The aim of the present experiment was determination 
of the $dp\rightarrow dp\eta$ cross sections near threshold 
in order to study the interaction
between the particles in the final state. 
The energy dependence of the total cross section is expected to
be very sensitive to the $d-\eta$ interaction. 
Even much weaker $p-\eta$ interaction significantly modifies this shape as observed in the
$pp\rightarrow pp\eta$ reaction\cite{Calen96,Calen99,Moskal04,hab}.

\section{Experiment}  
\begin{figure}[h]
\begin{center}
\parbox[c]{0.40\textwidth}{\includegraphics[width=0.4\textwidth]{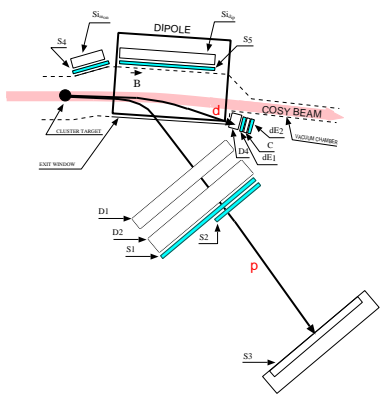}}
\parbox[c]{0.5\textwidth}{
\caption{
\label{detectionsystem}
Detection setup of the COSY-11 experiment\protect\cite{brau96}.
Protons from the $dp\rightarrow dp\eta$ reaction are measured in the drift chambers D1, D2,
as well as in the scintillator hodoscopes S1, S3. For detection of the outgoing deuterons,
small drift chamber D4 with hexagonal cells\protect\cite{smy05}, two scintillator detectors dE1,
dE2 and Cerenkov threshold counter were installed.
}
}
\end{center}
\end{figure}

\begin{figure}[h]
\begin{center}
\parbox[c]{0.40\textwidth}{\includegraphics[width=0.4\textwidth]{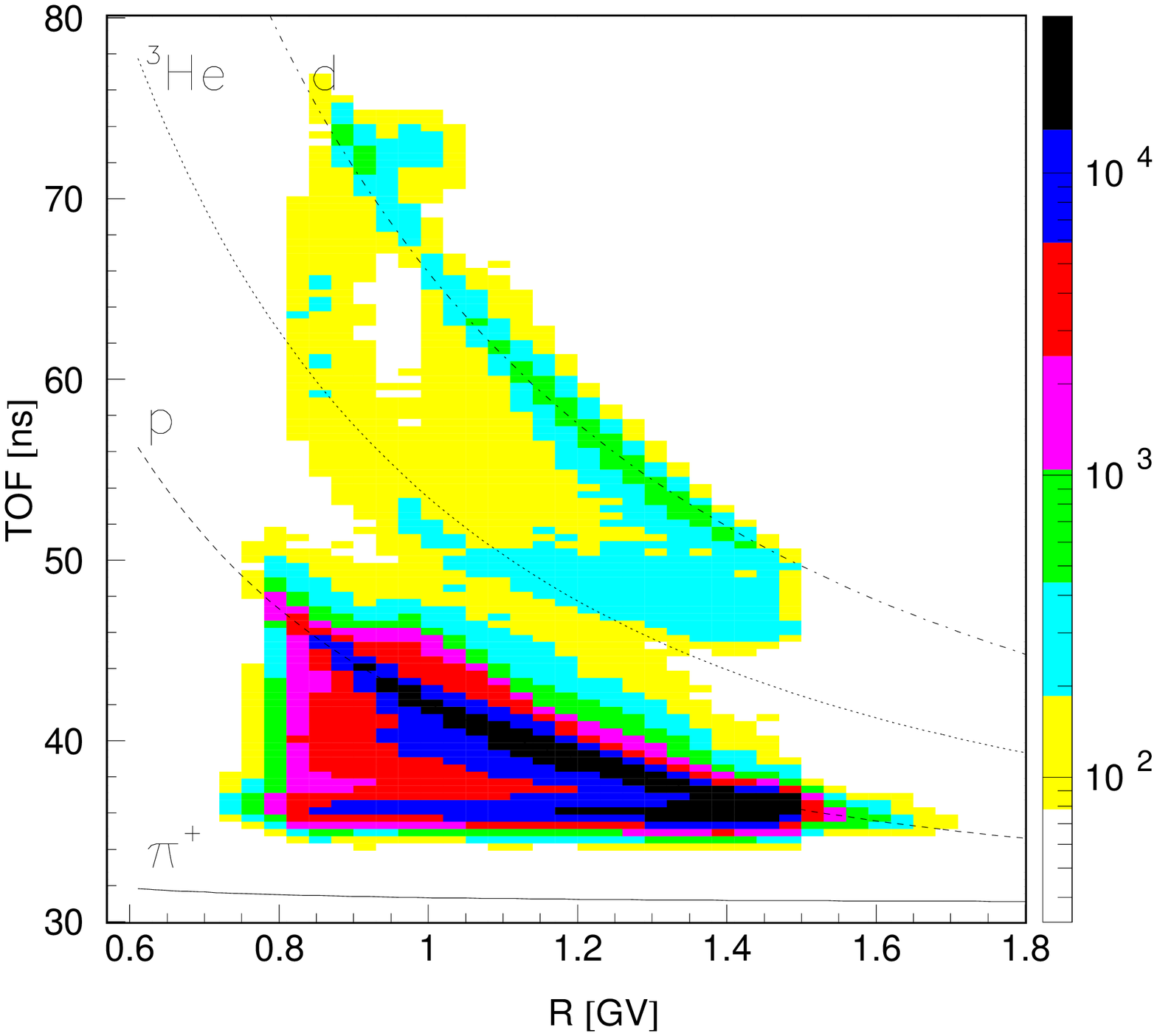}}
\parbox[c]{0.40\textwidth}{\includegraphics[width=0.4\textwidth]{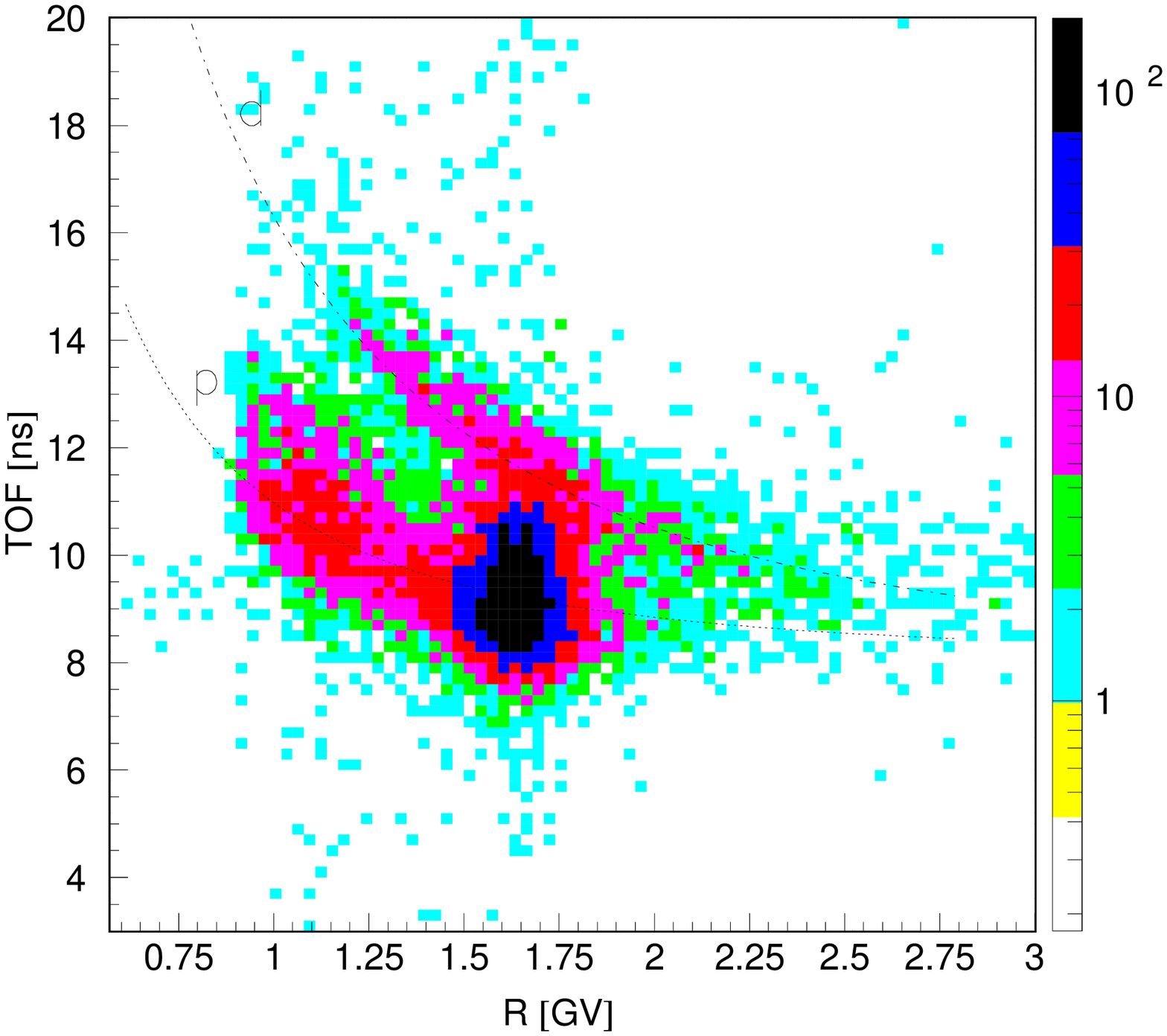}}
\caption{\label{tof_p_iden} \small Dependence of the TOF on magnetic rigidity
for particles registered in the drift chambers D1, D2 (left panel) and for particles 
detected in the D4 chamber (right panel).}
\end{center}
\end{figure}

The experiment was performed using the stochastically cooled deuteron beam 
of the COoler SYnchrotron (COSY) in Juelich\cite{maier97} 
scattered on internal proton target of the cluster jet type\cite{dombro97}.
The charged reaction products were registered with the COSY-11 detection system which is 
schematically shown in Fig.~\ref{detectionsystem}.
Protons from the $dp\rightarrow dp\eta$ reaction were measured in the drift chambers D1, D2 
and in the scintillator hodoscopes S1, S3. 
Tracking their trajectories through the magnetic field in the COSY-11 dipole magnet 
back to the target position allowed to determine their magnetic rigidity. 
Particle identification was based on the Time Of Flight (TOF) measured 
on the path of 9.3~m between the scintillator hodoscope S1 and S3. 
 Fig.~\ref{tof_p_iden} shows the TOF dependence on the magnetic rigidity
with clearly separated protons, deuterons and ${^3 \mbox{He}}$. 
Due to setting of the triggering electronics, only a tail of pion TOF distribution 
was registered as can be seen in the figure.
The outgoing deuterons from the $dp\rightarrow dp\eta$ reaction have about 
two times larger momenta than the outgoing 
protons and, consequently, their deflection in the magnetic field of the 
COSY-11 dipole magnet is two times smaller.
Their trajectories were measured with a small drift chamber D4, placed in the area between
the chamber D1 and the beam pipe, allowing for a three dimensional particle 
track reconstruction (see Fig.~\ref{detectionsystem}). 
Due to a huge background of spectator protons originating from the break-up
of the deuteron beam interacting with the proton target, identification
of the deuterons crossing the D4 chamber was a very difficult task.
Therefore, for the particle identification three different methods were applied 
and they were based on: (i) energy losses in two scintillation detectors dE1 and dE2 with a thickness
of 5~mm and 10~mm respectively, placed behind the D4 chamber, 
(ii) signal (or lack of signal) in a threshold Cerenkov detector made of 2~cm plexiglas plate
and (iii) TOF measured on the path of about 2.3~m between the target and the dE1 detector.  
For the TOF determination, the time of the reaction at the target was derived from measurement 
of velocity of the associated particle registered in the scintillator hodoscopes S1 and S3 
and from the time measured by the S1 detector. 
Dependence of the TOF on the magnetic rigidity for particles registered with the D4 drift chamber 
is shown in the right panel of Fig.~\ref{tof_p_iden}.
The $\eta$-mesons produced in the $dp\rightarrow dp\eta$ reaction were identified using the missing mass method.

The integral luminosity for each beam momentum was 
determined from acceptance corrected number of ${^3 \mbox{He}}\eta$ events measured simultaneously with
$dp\eta$ channel. The $dp\rightarrow {^3 \mbox{He}}\eta$ total cross section was taken from Ref.~\refcite{mayer96}.

Application of the stochastic cooling to the COSY deuteron beam guarantees a high quality of the beam
with the momentum smearing on the level of $\Delta$p/p $\approx$ 10$^{-4}$. 
However, the absolute beam momentum determined on the basis of the accelerator frequency 
is know with accuracy of $\Delta$p/p $\approx$ 10$^{-3}$ only.
Therefore, for a more precise beam momentum determination, which is crucial in the present
near threshold measurements, we used the center of mass momenta of the ${^3 \mbox{He}}$ ions 
from the $dp\rightarrow {^3 \mbox{He}}\eta$ reaction calculated from the analysis of ${^3 \mbox{He}}$
momenta in the magnetic field of the COSY-11 dipole magnet. 
Resulting correction to the nominal beam momentum was -1.6$\pm$0.4~MeV/c and the uncertainty of the
excess energy for the $dp \rightarrow dp\eta$ reaction was $\pm$0.1~MeV.  

\section{Preliminary results}
\begin{figure}[h]
\begin{center}
\parbox[c]{0.40\textwidth}{
\centering
\includegraphics[width=0.4\textwidth]{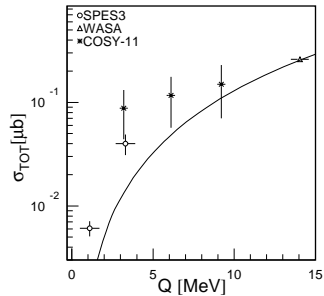}
}
\parbox{0.5\textwidth}{\caption{ Dependence of total cross section on the excess energy. The
line indicates three body phase-space normalized to the PROMICE/WASA point\protect\cite{bilg04}.
Open circles shows data from reference\protect\cite{Hibou00}, and stars denotes
preliminary results determined by the  COSY-11 group.
}}
\label{crossdpeta}
\end{center}
\end{figure}

The measurement was done for three deuteron beam momenta above the  $dp \rightarrow dp\eta$ threshold namely:
3177.4, 3189.4 and 3202.4~MeV/c and one momentum below the threshold equal to 3163.4~MeV/c.
The measurement below the threshold was used for the background subtraction under the eta peak in the missing
mass spectra. 
The determined $dp \rightarrow dp\eta$ total cross sections are shown in Fig.~\ref{crossdpeta} together
with the data measured at SATURNE and at WASA.
The data confirm a strong effect of the interaction in the $d p \eta$ meson final state.
The enhancement of the near-threshold cross sections with respect to the phase-space behavior
indicates for an effect of a strong interaction between the final state particles. We expect that 
further analysis of the collected experimental data will allow to reduce substantially the
relatively large uncertainties of our preliminary data points.

\section{Acknowledgments}
We acknowledge the support of the
European Community-Research Infrastructure Activity
under the FP6 "Structuring the European Research Area" program
(HadronPhysics and Hadron Physics-Activity -N4:EtaMesonNet contract number RII3-CT-2004-506078),
of the FFE grants (41266606 and 41266654) from the Research Center Juelich,
of the DAAD Exchange Programme (PPP-Polen),
of the Polish State Committee for Scientific Research
(grant No. PB1060/P03/2004/26).

\end{document}